\documentclass[
  preprint,
  linenumber,
  prl,
  aps,
  superscriptaddress,
  floatfix
]{revtex4-2}
\usepackage{graphicx}
\usepackage{dcolumn}
\usepackage{bm}
\usepackage{lineno}
\usepackage{amsmath}
\usepackage{comment}
\usepackage{enumitem}
\usepackage[dvipsnames]{xcolor}
\usepackage{float}
\usepackage{amsfonts}
\usepackage{dsfont}
\usepackage{bibunits}

\begin{document}

\preprint{} % submission number
%%%%%%%%%%%%%%%%%%%%%%%%%%%%%%%%%%%%%%
\title{Phason-driven temperature-dependent transport in moiré graphene}

%%%%%%%%%%%%%%%%%%%%%%%%%%%%%%%%%%%%%%
%% Affiliations
%%%%%%%%%%%%%%%%%%%%%%%%%%%%%%%
\author{Alex Boschi}
\affiliation{Center for Nanotechnology Innovation @NEST, Istituto Italiano di Tecnologia, I-56127, Pisa, Italy}
\author{Alejandro Ramos-Alonso}
\affiliation{Department of Physics, Columbia University, NY-10027, New York , USA}
\author{Vaidotas Mi\v{s}eikis}
\affiliation{Center for Nanotechnology Innovation @NEST, Istituto Italiano di Tecnologia, I-56127, Pisa, Italy}
\author{Kenji Watanabe}
\affiliation{Research Center for Electronic and Optical Materials, National Institute for Materials Science, 1-1 Namiki, Tsukuba, 305-0044, Japan}
\author{Takashi Taniguchi}
\affiliation{Research Center for Materials Nanoarchitectonics, National Institute for Materials Science, 1-1 Namiki, Tsukuba, 305-0044, Japan}
\author{Fabio Beltram}
\affiliation{NEST, Istituto Nanoscienze-CNR and Scuola Normale Superiore, Piazza San Silvestro 12, I-56127, Pisa, Italy}
\author{Stiven Forti}
\affiliation{Center for Nanotechnology Innovation @NEST, Istituto Italiano di Tecnologia, I-56127, Pisa, Italy}
\author{Antonio Rossi}
\affiliation{Center for Nanotechnology Innovation @NEST, Istituto Italiano di Tecnologia, I-56127, Pisa, Italy}
\author{Camilla Coletti}
\affiliation{Center for Nanotechnology Innovation @NEST, Istituto Italiano di Tecnologia, I-56127, Pisa, Italy}
\author{Rafael M. Fernandes}
\affiliation{Department of Physics and Anthony J. Leggett Institute for Condensed Matter Theory, The Grainger College of Engineering,
University of Illinois Urbana-Champaign,  IL 61801, Urbana, USA}
\author{Héctor Ochoa}
\affiliation{Department of Physics, Columbia University, NY-10027, New York , USA}
\author{Sergio Pezzini}
\email{sergio.pezzini@nano.cnr.it}
\affiliation{NEST, Istituto Nanoscienze-CNR and Scuola Normale Superiore, Piazza San Silvestro 12, I-56127, Pisa, Italy}

%%%%%%%%%%%%%%%%%%%%%%%%%%%%%%%%%%%%%%%
%% abstract
%%%%%%%%%%%%%%%%%%%%%%%%%%%%%%
\begin{abstract}
The electronic and vibrational properties of 2D materials are dramatically altered by the formation of a moiré superlattice. 
The lowest-energy phonon modes of the superlattice are two acoustic branches (called \textit{phasons}) that describe the sliding motion of one layer with respect to the other. Considering their low-energy dispersion and damping, these modes may act as a significant source of scattering for electrons in moiré materials. Here, we investigate temperature-dependent electrical transport in minimally twisted bilayer graphene, a moiré system developing multiple weakly-dispersive electronic bands and a reconstructed lattice structure. We measure a linear-in-temperature resistivity across the band manyfold above $T\sim{10}$ K, preceded by a quadratic temperature dependence. While the linear-in-temperature resistivity is up to two orders of magnitude larger than in monolayer graphene, it is reduced (approximately by a factor of three) with respect to magic-angle twisted bilayer graphene. Moreover, it is modulated by the recursive band filling, with minima located close to the full filling of each band.
Comparing our results with a semiclassical transport calculation, we show that the experimental trends are compatible with scattering processes mediated by longitudinal phasons, which dominate the resistivity over the contribution from conventional acoustic phonons of the monolayer. 
Our findings highlight the close relation between vibrational modes unique to moiré materials and carrier transport therein.
\end{abstract}

\maketitle

%%%%%%%%%%%%%%%%%%%%%%%%%%%%%%%%%%%%%%%
%% main part of the manuscript starts here
%%%%%%%%%%%%%%%%%%%%%%%%%%%%%%
\begin{bibunit}[apsrev4-2]
%% INTRO

Crafting moiré systems out of atomically thin materials has changed the paradigm of condensed-matter physics by providing a versatile platform to engineer, control, and investigate strongly correlated electronic phases \cite{andrei2021marvels} . Limiting our discussion to twisted bilayer graphene (TBG), the archetypic moiré system, key experimental observations at the so-called magic angle \cite{bistritzer2011moire} include correlated insulators \cite{cao2018correlated} , superconductors \cite{cao2018unconventional} , orbital magnets \cite{sharpe2019emergent} , quantum anomalous Hall insulators \cite{serlin2020intrinsic} , and cascades of phase transitions \cite{wong2020cascade, zondiner2020cascade} . Despite the variety of tuning knobs available, such as gating \cite{lu2019superconductors} , hydrostatic pressure \cite{yankowitz2019tuning} and proximity to screening layers  \cite{stepanov2020untying} or other 2D materials\cite{arora2020superconductivity}, temperature-dependent measurements of the resistivity $\rho(T)$, remain a primary tool to characterize the properties of TBG. Indeed, specific $\rho(T)$ trends have demonstrated phenomena such as the isospin Pomeranchuk effect \cite{saito2021isospin} and %Planckian dissipation in TBG \cite{cao2020strange, jaoui2022quantum} , the latter being signaled by 
an unusually persistent low-temperature linear-in-$T$ resistivity in the vicinity of correlated states \cite{cao2020strange, jaoui2022quantum}. Electrical transport, however, can be sensitive to several contributions, such as scattering from lattice vibrations, especially when employing temperatures of the order of the Bloch-Grüneisen scale \cite{polshyn2019large}.\\ Crucially, for twist angles $\theta\sim1^\circ$ and below, the lattice structure of TBG is largely reconstructed as a result of the interplay between interlayer (adhesion) and intralayer (elastic) forces \cite{yoo2019atomic} . 
%TBG thus forms a triangular array of commensurate AB and BA-stacked domains that minimize energy, separated by strain solitons, whereas unfavored AA-stacked regions shrinks at the vertices of the triangular areas. 
The resulting moiré pattern can be envisioned as a triangular lattice of shrunken AA-stacked areas (of maximum atomic overlap between graphene layers) connected by strain solitons separating domains of Bernal (AB and BA) stacking. Such reconstructed moiré pattern significantly alters the phonon spectrum. \cite{lian2019twisted, maity2020phonons, suri2021chiral, gao2022symmetry, samajdar2022moire, cappelluti2023flat, maity2023electrons, kang2025analytical, zhang2025atom, dantas2025raman} Specifically, a distinct class of low-energy collective excitations emerges from the interlayer shear motion. These moir\'e phonons describe fluctuations of the stacking order in the reconstructed superlattice and include two branches of acoustic-like modes known as phasons.\cite{ochoa2019moire} In the ideal, dissipationless limit, phasons can be understood as the Goldstone modes of the moiré incommensurate lattice describing the sliding motion of the strain solitons. 
However, in realistic systems, interlayer interactions, electronic coupling, and structural disorder introduce damping, rendering phasons overdamped and diffusive \cite{Ochoa2022}. This dissipative character is key to their influence on electronic transport, where phason scattering has been predicted to extend the linear-in-$T$ behavior of $\rho(T)$ down to temperatures lower than the Bloch-Grüneisen scale  \cite{ochoa2019moire, ochoa2023extended}. Moreover, recent results on the quantum twisting microscope (down to $\theta=5^\circ$) indicate that the electron-phason coupling in TBG increases with decreasing twist angle, while the energy of the modes decreases  \cite{birkbeck2025quantum}, suggesting that electrons might strongly couple to soft phasons at smaller $\theta$.\\
In this work, we investigate $\rho(T)$ of bilayer graphene at minimal twist angle  (mTBG, $\theta=0.36^\circ$). This regime has received comparatively less attention with respect to magic-angle TBG, mostly due to the absence of correlated electronic phases \cite{yoo2019atomic, lu2021multiple, shen2024strongly, nuckolls2025spectroscopy} . However, in the context of our study, mTBG permits to safely: (i) rule out the influence of electronic correlation on $\rho(T)$, and (ii) consider a reconstructed lattice structure with a triangular soliton network. These two conditions are common over a broad twist angle range \cite{yoo2019atomic, lu2021multiple, shen2024strongly}, making our results relevant beyond the specific angle investigated and robust to twist angle variations commonly observed in TBG devices \cite{uri2020mapping}. Combining gate and temperature-dependent transport experiments with theoretical modeling, we provide evidence for the prominent role of the electron-phason scattering in this system.
%% Results and discussion

\begin{figure}
    \includegraphics[width=0.5\linewidth]{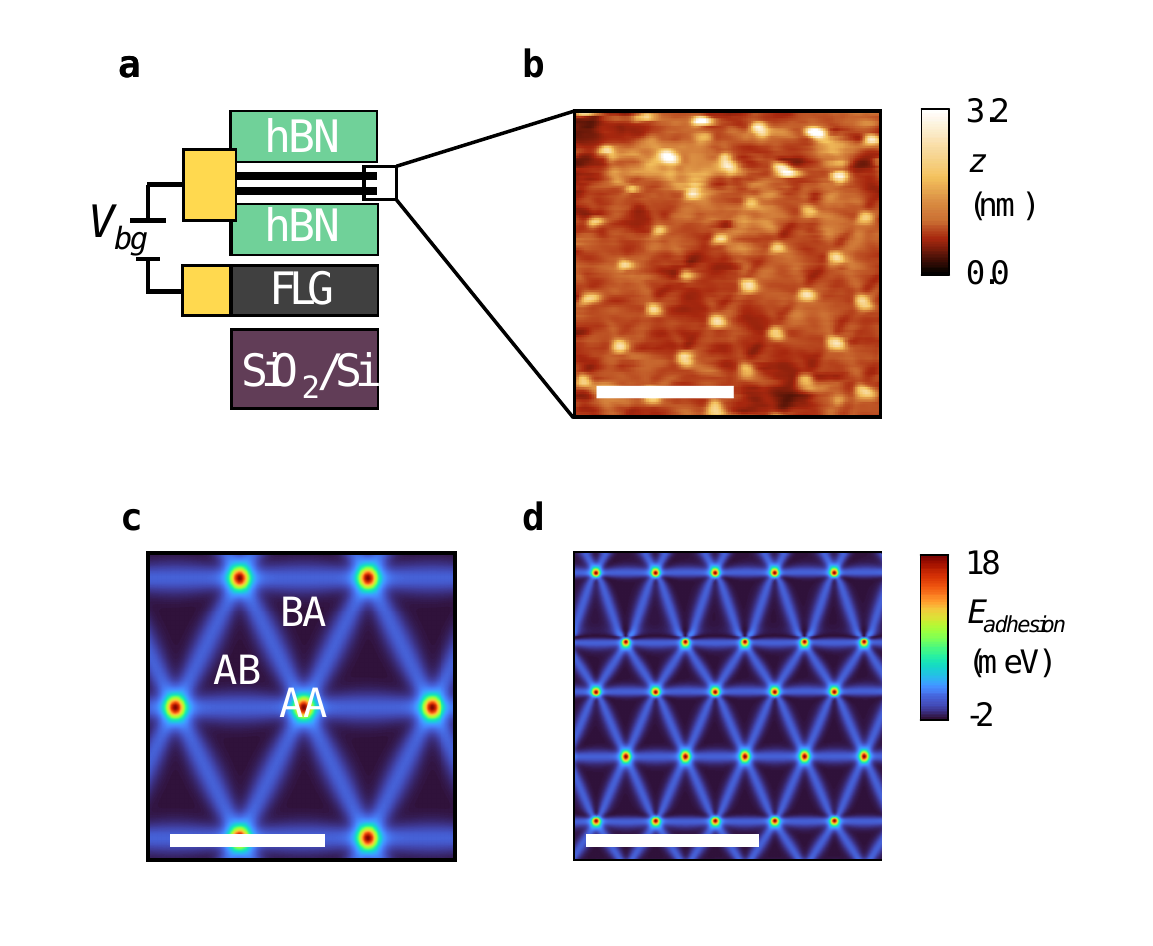}
  \caption{%Minimally twisted bilayer graphene forms a relaxed moiré pattern with phason modes. 
  (a) Schematics of the studied device. (b) STM topographic image of a mTBG sample with top exposed surface; sample temperature and tip current are 78 K and 0.4 nA respectively. (c) Adhesion energy density (per graphene's unit cell) between the two layers of mTBG ($\theta=0.4^{\circ}$) in real space after relaxation; the model parameters are the shear and bulk modulus of graphene, $\mu = 9.57 \text{ eV/\AA}^{2}$ and $B = 12.82 \text{ eV/\AA}^{2}$, respectively, and the adhesion energy constant, $V = 0.89$ meV/\AA$^{2}$ (see more details in the Supplemental Material \cite{supp}). 
  (d) Enlarged adhesion energy map corresponsing to a frozen longitudinal phason mode with wavevector $0.077$ nm$^{-1}$ along the $\Gamma-M$ axis of the moir\'e Brillouin zone. The color scale for the adhesion energy is the same in panel c and d. The scale bars are 100 nm in panels b, d, and 40 nm in c.}
  \label{fig1}
\end{figure}
A sketch of the studied device is depicted in Figure \ref{fig1}a. We employ a van der Waals (vdW) stack composed of mTBG sandwiched between two hexagonal Boron Nitride (hBN) flakes, with a few-layer graphite (FLG) flake acting as  back-gate electrode, deposited on a SiO$_{2}$/Si substrate. This structure follows the most recent strategies developed to minimize extrinsic scattering sources in vdW heterostructures \cite{rhodes2019disorder}. mTBG is artificially assembled from chemical vapor deposited (CVD) monolayer graphene single crystals, employing the grow-and-stack technique  introduced in Ref. \cite{piccinini2022moire} (an extension of the popular tear-and-stack\cite{kim2016van, cao2016superlattice} ; further details on  fabrication can be found in Supplemental Material \cite{supp}). \\
Figure \ref{fig1}b shows a scanning tunneling microscopy (STM) topographic map acquired on an equivalent mTBG free of the topmost hBN flake, which makes the moiré accessible to imaging. We observe a clear triangular pattern, with bright spots at vertices connected by straight domain walls. Comparable STM topographies of mTBG were recently reported in Ref. \cite{nuckolls2025spectroscopy} . The superlattice periodicity inferred from this acquisition is $\lambda=41$ nm, which corresponds to $\theta= 0.34^\circ$, approximating the twist angle of the mTBG transport device.\\
The STM pattern can be directly compared with the adhesion energy map of Figure \ref{fig1}c, calculated for a mTBG with comparable twist angle, and where we focus on a $\sim100\times100$ nm$^2$ area. The higher adhesion energy spots correspond to AA stacking, while the lower adhesion energy regions correspond to AB and BA stacking. In Figure \ref{fig1}d, we plot an enlarged adhesion energy map showing the longitudinal phason distortion of the moiré. 
These results are obtained by minimization of the mechanical free energy of the bilayer in a continuum model\cite{koshino2019moire, ochoa2025moirephonons}  describing the aforementioned competition between intralayer elastic forces and interlayer adhesion forces. 
The phason distortion corresponds to the lowest-energy harmonic fluctuation on top of the relaxed solution. Details of the model and its numerical implementation are discussed in the Supplemental Material \cite{supp}. 
%The methods employed to compute the adhesion energy maps are discussed in Supporting Information. \\
\begin{figure}
    \includegraphics[width=\linewidth]{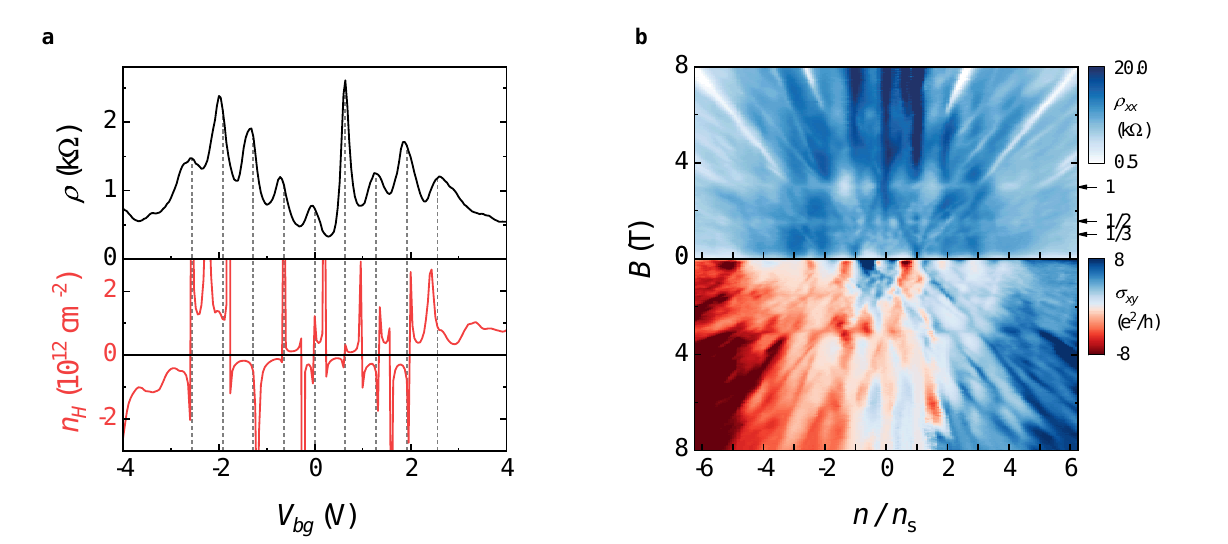}
  \caption{%Low-temperature electrical transport of multiple moiré bands and Hofstadter butterfly.
  (a) Resistivity of mTBG (black line, top panel) and Hall carrier density (red line, bottom panel) as a function of back-gate voltage. The Hall measurements are performed at a perpendicular magnetic  field $B=0.24$ T. The vertical dashed gray lines are located at integer $n/n_s$ values.  (b) Color maps of longitudinal resistivity (top) and Hall conductivity (bottom), as a function of band filling and magnetic field. The arrows indicate three representative conditions of rational values of flux quanta per superlattice unit cell. All data are acquired at $T=0.36$ K.}
  \label{fig2}
\end{figure}
We proceed with (magneto)transport measurements of mTBG at the lowest temperature available in our setups ($T = 0.36$ K). Figure \ref{fig2}a presents the zero-field resistivity $\rho$ (upper panel) and the Hall carrier density $n_H$ (lower panel) as functions of the voltage applied to the FLG back-gate $V_{bg}$. $\rho$ shows a series of evenly spaced peaks, consistent with previous reports on mTBG \cite{lu2021multiple, shen2024strongly} . Concomitantly with the peaks, we observe sign changes in $n_H$, signaling switching in the charge carrier type. This behavior is due to the successive filling of moiré bands, each of which accommodates four electrons (one per each spin and valley) or holes per superlattice unit cell. Considering the separation of the peaks ($V_{bg}=0.64$ V, see dashed gray vertical lines) and the gate lever arm ($0.5\times10^{12}$ cm$^{-2}$/V), we obtain the filling density of the moiré bands $n_s=0.32\times10^{12}$ cm$^{-2}$, which provides an estimate of the moiré periodicity ($\lambda=39$ nm) and the twist angle ($\theta=0.36^\circ$). Again similar to Refs. \cite{lu2021multiple, shen2024strongly} , we observe the strongest $\rho$ maximum at $n/n_s=1$, and a progressive dampening of the peaks for $\lvert n/n_s \rvert>3$. As discussed in the Supplemental Material \cite{supp}, several peaks show a weak insulating behavior even at zero magnetic field, suggesting that finite gaps separate the moiré bands. Additional sign changes in $n_H$, taking place approximately at half filling ($n/n_s\sim1/2, 3/2, ...$), signal Lifshitz transitions within each moiré band (see Theoretical Model in the Supplemental Material \cite{supp}) .\\
Under large perpendicular magnetic fields, we measure a clear Hofstader’s butterfly \cite{hofstadter1976energy, ponomarenko2013cloning, dean2013hofstadter,hunt2013massive} corroborating both the minimal twist angle and the electronic quality of the CVD-based mTBG sample. Figure \ref{fig2}b shows color maps of the longitudinal resistivity $\rho_{xx}$ and Hall conductivity $\sigma_{xy}$ as a function of band filling and magnetic field. At low field ($B<1$ T), we observe a series of fan-like structures originating from each integer value of $n/n_s$, confirming the sequential band filling controlled by $V_{bg}$. Increasing the magnetic field, commensurability conditions between the magnetic flux $\phi$ and the moiré are detected as horizontal lines located at rational values of flux quanta $\phi_0$ per superlattice unit cell (three of such conditions are highlighted in the upper panel of Figure \ref{fig2}b), where extended minima of $\rho_{xx}$ signal delocalized Brown-Zak fermions\cite{krishna2017high}. $\phi/\phi_0=1$ is reached at $B=3.1$ T, confirming the twist angle estimated above. For $\phi/\phi_0>1$, the data reveal a nontrivial unbound spectrum, as indicated by states at finite Chern number  ($C$, where $n/n_s=C\phi/4\phi_0$) interrupting $C=0$ (vertical) trivial states and connecting multiple bands \cite{lu2021multiple}. Comparable Hofstader's patterns are measured using different voltage probes in the device (see Supplemental Material \cite{supp}).\\
\begin{figure}
    \includegraphics[width=\linewidth]{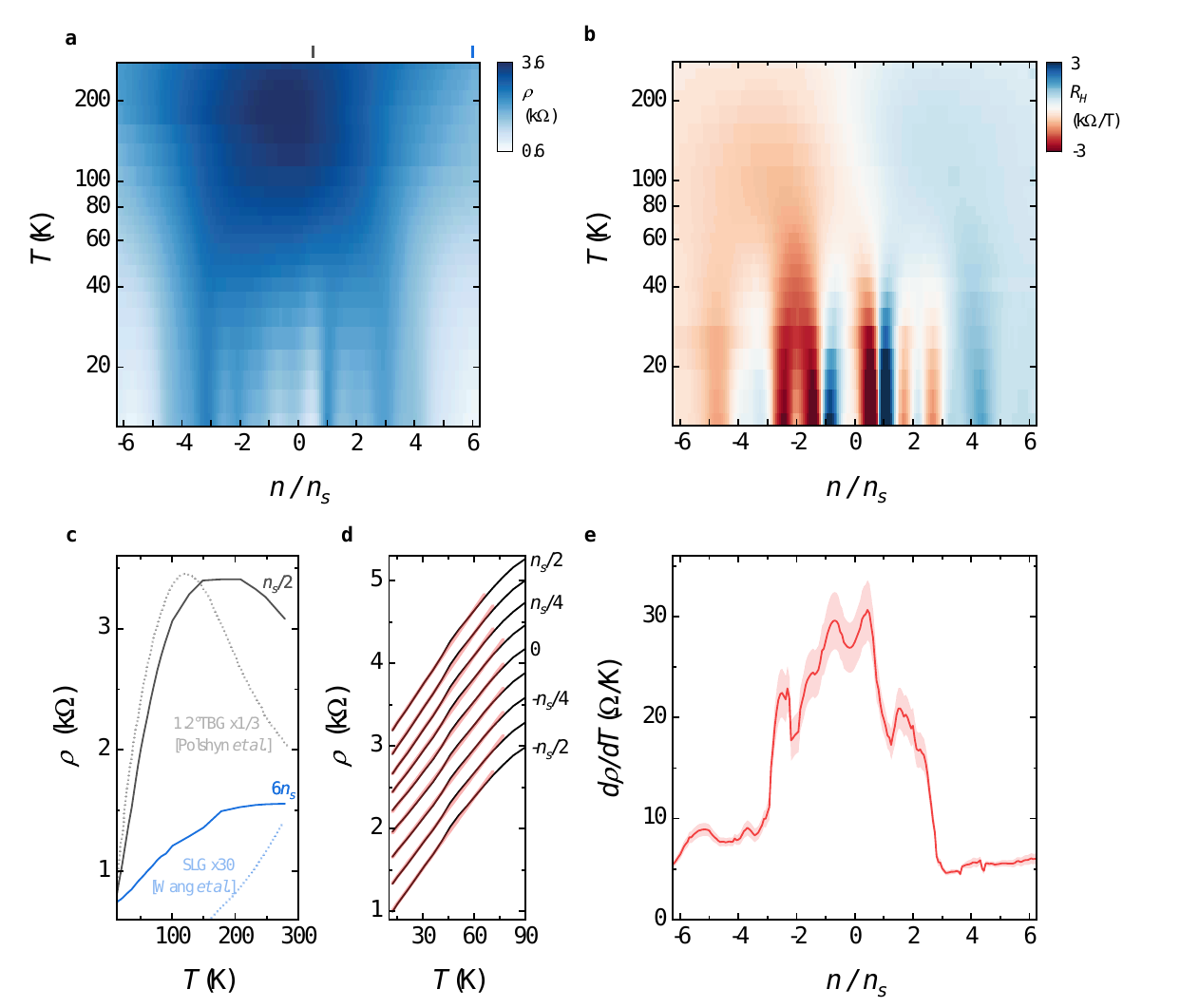}
  \caption{%Thermal activation across moiré bands at intermediate-to-high temperatures and linear-in-\textit{T} resistivty. 
  (a) Resistivity and (b) Hall coefficient $R_H$ (measured at $B=0.25$ T) as a function of band filling and temperature (Log scale). (c) $\rho(T)$ at two representative values of band filling (see markers in panel a). Literature reference curves for 1.2°-TBG (gray dotted line) and monolayer graphene (MLG, light blue dotted line) are reported for comparison from Ref. \cite{polshyn2019large} and Ref. \cite{wang2013one} , respectively. (d) $\rho(T)$ at selected fillings within the first moiré band (black lines). The curves are offset by 0.3 k$\Omega$ for clarity. The solid red lines are linear fits. (e) Extracted $d\rho/dT$ from linear fits as a function of band filling. The shaded red area corresponds to $\pm$ one standard error from the fits.}
  \label{fig3}
\end{figure}
We now focus on the temperature-dependent electrical transport of mTBG, starting from an intermediate-to-high $T$ range (12 K $<T<$ 300 K). Figure \ref{fig3}a shows a color map of $\rho$ as a function of band filling and temperature, where one can identify several regimes. At large band filling ($\lvert n/n_s \rvert>3$), $\rho(T)$ is essentially featureless and grows monotonically with temperature. A richer behavior is found for $\lvert n/n_s \rvert<3$, where resistivity peaks at full band filling are visible up to $\sim80$ K. Above this temperature, the peaks collapse in a single broad maximum (see Supplemental Material \cite{supp}) with saturated $\rho(T)$, which finally decreases at the highest temperatures ($T>200$ K). The behavior of $\rho(T)$ correlates with the Hall coefficient $R_H$, shown in Figure \ref{fig3}b. Here, repeated sign changes within and between the moiré bands disappear around the same temperature at which $\rho(T)$ saturates. A similar scenario was reported in Ref.\cite{polshyn2019large} for near-magic-angle TBG and attributed to thermal activation from higher-energy dispersive bands. The saturation of $\rho(T)$ (smearing of $R_H$) thus determines a threshold up to which transport can be considered as limited to individual low-energy  bands. We note that the saturation region forms a dome-like shape, indicating that bands at higher energies are progressively more affected by thermal activation from the dispersive compartment. The energy scale of the saturation temperature is in line with the bandwidth measured by scanning tunneling spectroscopy in Ref. \cite{nuckolls2025spectroscopy} ($<20$ meV for each moiré band).\\
In Figure \ref{fig3}c we show $\rho(T)$ at two representative band fillings ($n_s/2$ and $6n_s$). For reference, we show on the same plot digitalized literature data of $\rho(T)$ for $1.2^\circ$-TBG and monolayer graphene, at comparable carrier density, respectively from Ref.\cite{polshyn2019large}  and Ref.\cite{wang2013one} . The behavior of mTBG at low filling closely resembles that of near-magic-angle TBG, up to an approximate scaling factor of 1/3 and a weaker downturn after saturation. At large band filling, $\rho(T)$ is dampened; however, it is still larger than in monolayer graphene by more than one order of magnitude, indicating the key role played by the moiré superlattice in the observed phenomenology. At intermediate temperatures (below the saturation threshold), both curves from mTBG show a linear-in-\textit{T} behavior but with largely different magnitudes. The persistency of the linear-in-\textit{T} trend is exemplified in Figure \ref{fig3}d, where we plot several $\rho(T)$ curves within the two lowest-energy bands (black lines), along with linear fits (superimposed red lines, see  Supplemental Material \cite{supp} for details on the fitting procedure). This behavior extends across multiple moiré bands (as shown in Supplemental Material \cite{supp}), and over the whole carrier density range span in the experiment, allowing to extract the coefficient $d\rho/dT$ shown in Figure \ref{fig3}e. $d\rho/dT$ is strongly modulated as a function of band filling, with minima mapping approximately integer fillings, while maxima are located around half-fillings. While minima at full filling are consistent with the behavior at $n=0$ reported in Ref. \cite{polshyn2019large} , we note that the maxima close to half filling are not observed at any of the twist angles investigated therein. $d\rho/dT$ progressively decreases with band filling, until it is abruptly suppressed for $|n/n_s|>3$, indicating that the driving mechanism of the large linear-in-$T$ resistivity is restricted to the lower-energy moiré bands. The same behavior is reproduced using different voltage probes (see Supplemental Material \cite{supp}). \\
\begin{figure}
    \includegraphics[width=0.77\linewidth]{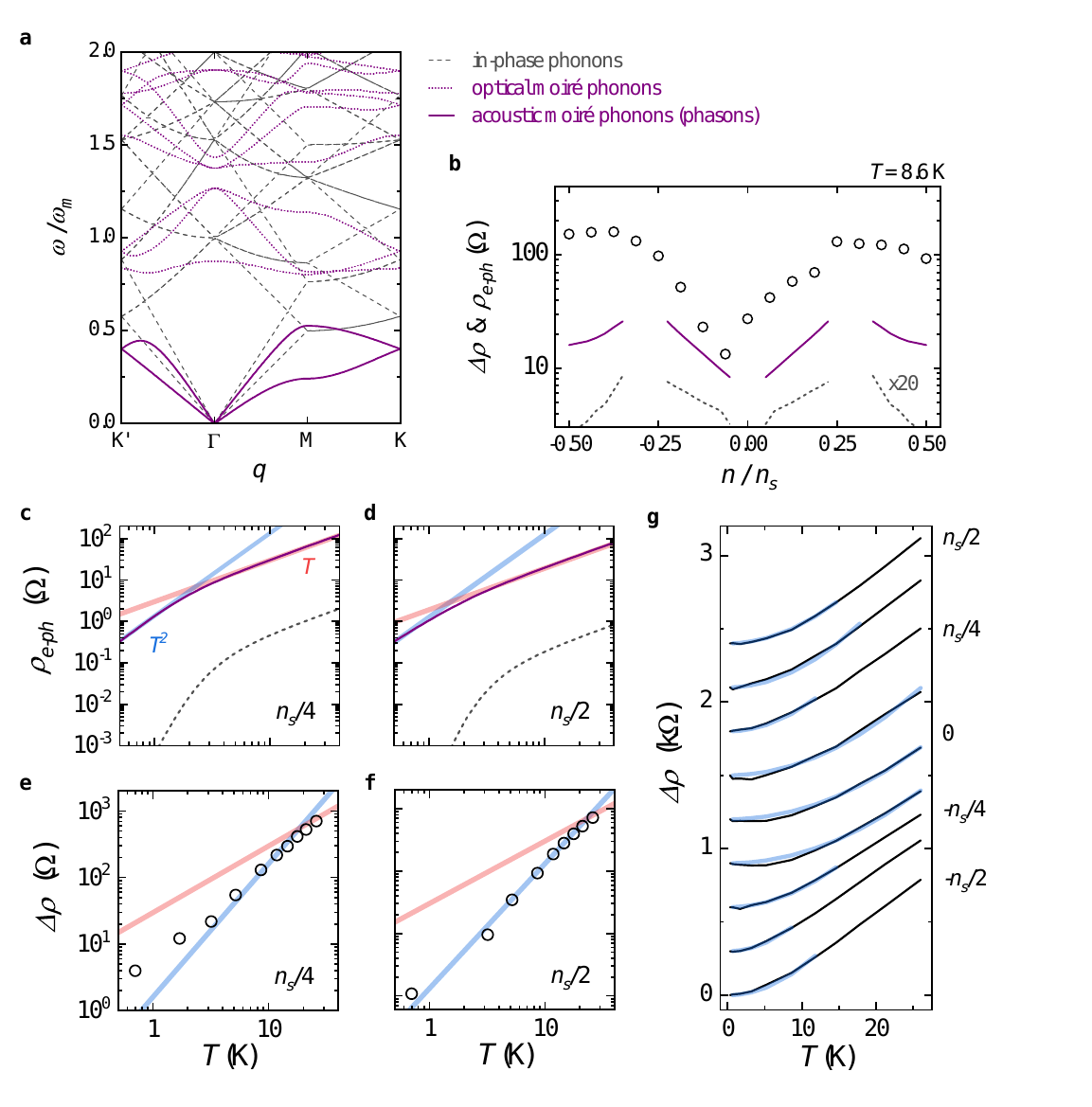}
  \caption{
  (a) Phonon bands of mTBG within the moiré Brillouin zone. The line styles and colors identify three groups of vibrational modes: in-phase phonons (gray dashed), optical moiré phonons (purple dotted), and phasons (purple continuous). The same code applies to contributions of these modes to $\rho_{e-ph}$ plotted in panels b, c and d.  (b) Experimental \textit{T}-dependent resistivity $\Delta\rho$ (open black circles) and computed resistivity $\rho_{e-ph}$ for carriers scattering off of in-phase phonons and phasons, as a function of band filling, at $T=8.6$ K. The model parameters are reported in the Supplemental Material \cite{supp}. (c,d) Computed $\rho_{e-ph}$ for the longitudinal phason (solid) and transverse in-phase phonon (dashed) modes, as a function of $T$ at $n_s/4$ and $n_s/2$. The light blue and red lines are guides to the eye for the quadratic and linear-in-$T$ trends of the phason contribution. (e,f) $\Delta\rho$ as a function $T$, at $n_s/4$ and $n_s/2$. The light blue and red lines are quadratic and linear fits to the data, respectively, performed over different $T$ ranges. The y axis is the same in panels c-d, as also in panels e-f; the x axis is the same in panels c, d, e and f. (g) $\Delta\rho$ as a function of temperature at selected fillings within the first electronic band. The curves are offset by 0.3 k$\Omega$ for clarity.  The light blue lines are quadratic fits to the data.}
  \label{fig4}
\end{figure}
To elucidate the contributions of lattice vibrations to $\rho(T)$ in mTBG, we contrast the data with theoretical calculations at $\theta=0.4^\circ$.  
In Figure \ref{fig4}a, we plot the spectrum of in-phase monolayer acoustic phonons (gray dashed lines, corresponding to the center-of-mass vibrations of the bilayer), which are insensitive to the moir\'e pattern, and the calculated spectrum of moir\'e phonons (purple lines) describing out-of-phase (i.e., relative) long-wavelength vibrations of the bilayer, whose energy spectrum is reconstructed by the adhesion potential. The phason branches are shown as solid  lines. The frequency axis is normalized to the characteristic energy scale, which is inversely proportional to $\lambda$, corresponding to $\omega_{m}=1.92$ meV for this twist angle.  
For the electronic degrees of freedom, we consider the continuum model of TBG\cite{castroneto2007model, bistritzer2011moire, koshino2019electrons,balents2019model} with the interlayer tunneling parameters corrected by relaxation effects as in Ref.\cite{lu2021multiple} . The calculations correctly capture the position of the Lifshitz transition within the first moiré band at $n/n_s=\pm0.3$ (see details in the Supplemental Material \cite{supp}). However, we obtain partially overlapping bands, not reproducing the small gaps found experimentally, hence we limit our theoretical analysis to low carrier concentrations, $|n/n_s|<0.5$. We then follow a similar approach as Ref. \cite{ochoa2023extended} and compute the resistivity from a variational solution of the Boltzmann equation\cite{ziman2001book} using the explicit matrix elements of the electron-phonon and electron-phason couplings deduced from the continuum model (see details in the Supplemental Material \cite{supp}).
In both cases longitudinal modes dominate the resistivity over the rest (note that the longitudinal phason is in fact the lowest energy acoustic mode).\\
The computed $\rho_{e-ph}$ at $T=8.6$ K as a function of band filling is plotted in Figure \ref{fig4}b as solid purple lines (dotted grey lines) for the longitudinal phason (in-phase phonon, amplified by a factor of 20). 
In the same panel, we plot the experimental data at $T=8.6$ K, from which we subtracted the resistivity at base temperature (0.36 K) to isolate the temperature-dependent contribution $\Delta\rho$ (open black circles). The calculated $\rho_{e-ph}$ terms follow the same behavior with band filling as measured experimentally, with an increase in resistivity from $n/n_s=0$ to the Lifshitz transition ($n/n_s=\pm0.3$) and a subsequent decrease. In our model, both phonons and phasons generically couple to electrons on each graphene layer through a pseudo-gauge and a scalar (deformation) potential. Additionally, phasons couple to interlayer tunneling events due to the modulation of the moir\'e pattern. Accounting for the screening of the deformation potential in the Thomas-Fermi approximation, we find that the phason contribution to the resistivity dominates over the phonon contribution by at least one order of magnitude. \\
To further discriminate between conventional acoustic phonons and phasons, we focus on the temperature dependence of the corresponding $\rho_{e-ph}$ contributions, as shown in the calculations of Figure \ref{fig4}c and \ref{fig4}d at two representative values of band filling. 
In these and previous calculations, the spectral function of phason mode $\alpha$ entering in the scattering rates of electrons is given by\begin{align}
    \chi_{\alpha}''(\boldsymbol{k},\omega) = \frac{2}{\varrho} \frac{\gamma \omega}{[\omega^{2}-\omega^{2}_{\alpha}(\boldsymbol{k})]^{2}+\gamma^{2}\omega^{2}},
\end{align}
where $\varrho$ is graphene's mass density, $\omega_{\alpha}(\boldsymbol{k})$ is the phason frequency, and $\gamma$ is a phenomenological damping constant (see Ref. \cite{ochoa2023extended}). A similar expression applies to in-phase phonons, but in that case $\gamma\rightarrow 0$ as $\boldsymbol{k}\rightarrow 0$, as prescribed by Goldstone theorem, so these modes always remain propagating waves. However, in the case of phasons, the relative momentum of the layers is not a conserved quantity and, therefore, $\gamma\neq 0$ in general. In our calculations we choose $\gamma$ close to the critical damping condition, specifically, $T_{\gamma}=0.6T_{\textrm{BG}}$, where $T_{\textrm{BG}}$ is the Bloch-Grüneisen temperature, and $T_{\gamma}\equiv \hbar\gamma/k_B$ is the new scale introduced by the finite damping. Observe that the Bloch-Grüneisen scale is defined by the maximum momentum transferred within the Fermi surface and depends on the electron filling. However, for fillings beyond $|n/n_s|=0.15$ we find that its value saturates to $T_{\textrm{BG}}\sim9$ K and $T_{\textrm{BG}}\sim13$ K for phasons and in-phase phonons, respectively.\\
Both types of collective modes give  a linear-in-$T$ resistivity for temperatures above the Bloch-Grüneisen scale. 
However, the quadratic in $T$ resistivity at the lowest temperatures is a characteristic feature of phason scattering, since those processes are dominated by long-wavelength incoherent fluctuations,\cite{ochoa2023extended} while scattering by underdamped phonons gives rise to higher powers (typically $T^4$ in 2D) in the so-called Bloch-Grüneisen regime.
In Figure \ref{fig4}e-f, for comparison, we plot the experimental data $\Delta\rho(T)$ over the same temperature range used in the calculations in Figure \ref{fig4}c-d. We find that the low-temperature part follows a quadratic trend (blue fitting line), before crossing over to the linear-in-$T$ behavior discussed previously (red fitting lines). 
Multiple low-temperature $\Delta\rho(T)$ curves measured within the first moiré band (black lines) are presented in Figure \ref{fig4}g, together with $T^2$ fits (blue lines), showing that this behavior is consistently found at different band fillings (see \cite{supp} for additional curves across the different bands). While a $T^2$ resistivity is often associated with electron-electron scattering combined with umklapp processes \cite{baber1937contribution}, the fact that we find this behavior even at $n=0$, where the charge carrier density vanishes, shows that the $T^2$ behavior cannot be attributed to this mechanism. Moreover, although a $T^2$ behavior due to electron-hole friction was reported in charge-neutral TBG in Ref. \cite{bandurin2022interlayer} , we note that this mechanism requires charge compensation induced via a displacement electric field, which we can exclude in our single-gated device. Even considering compensation between thermally-activated electrons and holes, the observed linear-in-$T$ resistivity at intermediate temperatures leads us to exclude the friction picture. Thus, we conclude that the $T^2$ behavior of the resistivity arises from electron-phason scattering. Currently, our investigation  falls short of matching the exact magnitude of $\rho(T)$ and the exact location of the $T^2$ to linear-in-$T$ crossover. Further calculations including Hartree-Fock corrections to the electronic bands, as well as experiments at different twist angle, might shed additional light on this issue.\\
%% Conclusions
In conclusion, our combined experimental and theoretical investigation points to phasons as the dominant source of electron scattering in mTBG, revealing their central role in shaping temperature-dependent resistivity in the absence of electronic correlations. These findings underscore how the reconstructed moiré superlattice and its associated low-energy collective excitations can fundamentally influence charge transport. Beyond graphene, this work contributes to a growing understanding of phasons as a general mechanism in moiré materials. For instance, phasons have been proposed as a crucial mechanism enabling long-range exciton diffusion in heterostructures of transition metal dichalcogenides at low temperatures, where conventional phonon-assisted processes are otherwise suppressed\cite{maity2020phonons, rossi2024anomalous}. Our results thus offer new insights into the electron–lattice interplay in vdW systems and open perspectives for tuning transport and optical properties through structural engineering of moiré superlattices.

\section*{Acknowledgments}
A.B., V.M. and C.C. acknowledge financial support from the European Union through the GraPh-X project, under the grant agreement 101070482. F.B., S.F., A. R., C.C. and S.P. acknowledge financial support from PNRR MUR project No. PE00000023 - NQSTI. K.W. and T.T. were supported by the JSPS KAKENHI (Grant Numbers 21H05233 and 23H02052), the CREST (JPMJCR24A5), JST and World Premier International Research Center Initiative (WPI), MEXT, Japan, for hexagonal boron nitride crystals growth.

\end{bibunit}

\clearpage
%%%%%%%%%%%%%%%%%%%%%%%%%%%%%%%%%%%%%%%
%% Supplemental Material
%%%%%%%%%%%%%%%%%%%%%%%%%%%%%%
%%%%%%%%%%%%%%%%%%%%%%%%%%%%%%%%%%%%%%
%% SM included in Main
%%%%%%%%%%%%%%%%%%%%%%%%%%%%%%%
\newcommand*\mycommand[1]{\texttt{\emph{#1}}}
\setcounter{figure}{0}
\renewcommand{\thefigure}{S\arabic{figure}}
\setcounter{equation}{0}

%DEFINITIONS
\def\bra#1{\mathinner{\langle{#1}|}}
\def\ket#1{\mathinner{|{#1}\rangle}}
\newcommand{\1}{\mathds{1}}

%%%%%%%%%%%%%%%%%%%%%%%%%%%%%%%%%%%%%%%%%%%%%%%%%%%%%%%%%%%%%%%%%%%%%
\begin{bibunit}[apsrev4-2]
\section{Supplemental Material for \\ Phason-driven temperature-dependent transport in moiré graphene}
%%%%%%%%%%%%%%%%%%%%%%%%%%%%%%%%%%%%%%%
\subsection{Methods for device fabrication and transport measurements}

Monolayer graphene single crystals are synthesized by low-pressure chemical vapor deposition on copper foil and transferred on SiO$_2$/Si substrate following protocols described in Ref. \cite{giambra2021wafer} . Graphite and hBN are micro-mechanically exfoliated from bulk crystals with adhesive tape on SiO$_2$/Si.
The studied vdW stack is assembled via dry pick-up employing a polydymethilsiloxane (PDMS) dome coated with a poly-carbonate bisphenol-A  (PC) membrane \cite{zomer2014fast, purdie2018cleaning}. First, a graphite/hBN gate stack is released on a SiO$_2$/Si substrate with pre-defined Au alignment markers. After dissolving the PC membrane, the exposed hBN surface is cleaned of polymer residues using contact-mode atomic force microscopy (AFM) \cite{goossens2012mechanical}. Then, a second PDMS/PC stamp is used to pick-up sequentially an hBN flake and a graphene monolayer crystals. The transfer stage is then rotated by $\sim 1^\circ$ before picking-up a second graphene crystal, crystallographically aligned to the first one thanks to growth on a single Cu grain \cite{piccinini2022moire}. The minimal twist angle is set by relaxation of TBG during vdW assembly, as reported in Ref. \cite{lu2021multiple} . The hBN/mTBG stack is then released on top of the graphite/hBN gate stack.
Edge contacts \cite{wang2013one} to mTBG are defined using e-beam lithography, CF$_4$/O$_2$ reactive ion etching and thermal evaporation of Cr/Au (5/60 nm). E-beam lithography and CF$_4$/O$_2$ reactive ion etching are repeated to pattern a Hall bar mesa over bubble-free areas identified via optical microscopy (an optical image of the  device is shown in Figure \ref{figSI_transpHof}a). \\
The exposed mTBG used for STM characterization was assembled via a pick-and-flip method \cite{wong2020cascade, montanaro2023sub} involving a polyvinyl alcohol (PVA) membrane on top of the PDMS/PC stamp. hBN/mTBG is assembled on PDMS/PC/PVA and released on a second PDMS/PC stamp by dissolving PVA in water, leading to flipping of the stack. The flipped structure is deposited on a pre-patterned Cr/Au  pad for electrical grounding and  mechanically cleaned by contact-mode AFM. The sample is annealed in ultra-high vacuum at 170$^\circ$C for 6 hours and at 350$^\circ$C for 2 hours before performing STM measurements.\\
Magnetotransport measurements from 0.36 K to 26 K are performed in a dry “ICE 300 mK He-3” cryostat. A second set of measurements (12 K to 300 K) is acquired in a separate cool-down in a dry “ICE 3 K INV” cryostat. In both set-ups, four-probe measurements are performed with low-frequency (13 Hz) lock-in detection with constant current excitation (10-100 nA). The  graphite back-gate is biased with a dc source-meter. A constant voltage (-40 V) is applied to the Si substrate using a second dc source-meter to dope the contact regions outside of the graphite back-gate. Hall effect measurements are performed at opposite directions of magnetic field and anti-symmetrized.

\subsection{Theoretical model}

\subsubsection{Lattice relaxation and moir\'e phonons.}
We use continuum models to study the lattice and electronic degrees of freedom of TBG.\cite{ochoa2019moire} The models are formulated in terms of a vector-valued field $\boldsymbol{\phi}(\boldsymbol{r})$ describing the local stacking configuration around a lateral position $\boldsymbol{r}$ of the bilayer (in Eulerian coordinates) defined by a relative translation $\boldsymbol{\phi}$ of the layers starting from maximum lattice overlap (AA stacking). By definition, $\boldsymbol{\phi}(\mathbf{r})$ and $\boldsymbol{\phi}(\mathbf{r})+\mathbf{R}$ parametrize the same stacking configuration, where $\mathbf{R}$ is a Bravais vector of the graphene lattice.

We assume that these fields are smooth on the scale of graphene's lattice constant, $a=2.47$ \AA; this is guaranteed for small twist angles provided that the interlayer adhesion forces are small compared to the stiff lateral response of graphene. In order to produce the adhesion energy maps and moir\'e phonon dispersions shown in the main text, we start from a elasticity model of each monolayer. 
The effects of the moir\'e superlattice are introduced via an adhesion potential term. 
The resulting free energy reads
\begin{align}\label{eqn:free energy}
    \mathcal{F}[\boldsymbol{\phi}] = 
    \int d\boldsymbol{r} \left[ \frac{B-\mu}{4}(\partial_{i}\phi_{i})^{2} + \frac{\mu}{8}(\partial_{i}\phi_{j} + \partial_{j}\phi_{i})^{2}
    + V\sum_{n=1}^{3}\cos(\boldsymbol{g}_{n}\boldsymbol{\phi}(\boldsymbol{r}))\right].
\end{align}
Repeated indices are summed over, and the vectors $\boldsymbol{g}_{n}$ are graphene reciprocal lattice vectors in the first star related by $C_{3}$ rotations. 
The shear and bulk moduli of graphene are\cite{koshino2019moire} $\mu = 9.57 \text{ eV/\AA}^{2}$ and $B = 12.82 \text{ eV/\AA}^{2}$. The only parameter of the adhesion term, $V = 0.89$ meV/\AA$^{2}$, is related to the energy difference between AA and AB/BA (Bernal) stacking configurations. Note that $D_6$ symmetry forbids the presence of odd terms in the potential.

To calculate the relaxed configuration of the stacking field and the moir\'e phonon dispersions and polarizations we follow the method in Ref. \cite{ochoa2025moirephonons} . We first determine the stacking texture $\boldsymbol{\phi}_{0}(\boldsymbol{r})$ that minimizes the free energy in Eq.~\eqref{eqn:free energy} subjected to an imposed global twist angle $\theta$. For that, we write \begin{align}
\label{eq:eq_texture}
\boldsymbol{\phi}_{0}(\boldsymbol{r})=2\sin\left(\frac{\theta}{2}\right)\boldsymbol{\hat{z}}\times\boldsymbol{r}+\boldsymbol{u}_0(\boldsymbol{r}).
\end{align}
The first term corresponds to the stacking texture produced by a rigid rotation, while the second term represents the heterostrain field produced by lattice relaxation. The latter admits a Fourier decomposition in moir\'e harmonics, \begin{align}
\boldsymbol{u}_0(\boldsymbol{r})=\sum_{\boldsymbol{G}} e^{i\boldsymbol{G}\cdot\boldsymbol{r}}\,\boldsymbol{u}_{\boldsymbol{G}}.
\end{align}
The coefficients $\boldsymbol{u}_{\boldsymbol{G}}$ are then determined self-consistently by solving numerically the saddle-point equations derived from the functional in Eq.~\eqref{eqn:free energy} in an iterative process. The energy maps in the main text are the result of evaluating the adhesion potential at $\boldsymbol{\phi}_{0}(\boldsymbol{r})$ (multiplied by graphene's unit cell area).%, 5.28 \AA$^{2}$.

Next, we write
\begin{align}
    \boldsymbol{\phi}(\boldsymbol{r},t) = \boldsymbol{\phi}_{0}(\boldsymbol{r}) + \delta\boldsymbol{\phi}(\boldsymbol{r},t),
\end{align}
where we have decomposed the field into the relaxed stacking configuration and small dynamical fluctuations ($|\delta\boldsymbol{\phi}|<a$) due to long-wavelength layer-shear vibrations (i.e., moir\'e phonons). To determine their dispersion, we exploit the moir\'e translational symmetry of the problem by introducing Fourier series of the form \begin{align}
    \label{eq:fluctuation_Fourier}
\delta\boldsymbol{\phi}(\boldsymbol{r},t) 
    = \frac{1}{\sqrt{A}}\sum_{\boldsymbol{q}\in\textrm{mBZ}}\sum_{\boldsymbol{G}} \delta\boldsymbol{\phi}_{\boldsymbol{q}+\boldsymbol{G}}(t) e^{i(\boldsymbol{q}+\boldsymbol{G})\boldsymbol{r}}.
\end{align}
Hereafter $A$ denotes the area of the bilayer. The harmonic expansion of the functional in Eq.~\eqref{eqn:free energy} defines a dynamical matrix for each $\boldsymbol{q}$ restricted to the first Brillouin zone of the moir\'e (mBZ). We numerically diagonalize this matrix to obtain the moir\'e phonon dispersions and eigenvectors\cite{ochoa2025moirephonons}. The phonon frequencies in the main text are represented in units of the characteristic scale imposed by folding, \begin{align}
    \omega_{\textrm{m}}=\sqrt{\frac{\mu}{3\varrho}}\frac{4\pi}{\lambda},
\end{align}
where $\varrho = 7.55\cdot 10^{-7}$ kg/m$^{2}$ is the graphene's mass density, and $\mu = 9.57 \text{ eV/\AA}^{2}$ and $\lambda = 3.25 \text{ eV/\AA}^{2}$ are the Lamé parameters of graphene.

Finally, note that the in-phase phonons follow trivially from the same theory by taking $V=0$ in the previous equations, so their dispersion is simply the original linear acoustic phonons of graphene folded onto the mBZ.

\subsubsection{Electronic Hamiltonian}

The electronic Hamiltonian in second quantization is given by
\begin{align}
    \hat{H}[\boldsymbol{\phi}(\boldsymbol{r})] =
    \sum_{\xi=\pm 1}\int d\boldsymbol{r}\: \hat{\psi}^{\dagger}_{\xi}\: \hat{\mathcal{H}}^{(\xi)}[\boldsymbol{\phi}(\boldsymbol{r})] \:\hat{\psi}_{\xi},
    %\:\text{ with }\:
   % \hat{\mathcal{H}}^{(\xi)}[\boldsymbol{\phi}(\boldsymbol{r})] = 
  %  \begin{pmatrix}
 %   \hat{\mathcal{H}}^{(\xi)}_{t}[\boldsymbol{\phi}(\boldsymbol{r})] & \hat{\mathcal{T}}^{(\xi)}[\boldsymbol{\phi}(\boldsymbol{r})] \\ \hat{\mathcal{T}}^{(\xi)\dagger}[\boldsymbol{\phi}(\boldsymbol{r})] & \hat{\mathcal{H}}^{(\xi)}_{b}[\boldsymbol{\phi}(\boldsymbol{r})]
%    \end{pmatrix}.
\end{align}
where $\hat{\psi}_{\xi}$ are 4-component spinor field operators representing electronic excitations around graphene's valleys $\xi=\pm1$, and $\hat{\mathcal{H}}^{(\xi)}[\boldsymbol{\phi}(\boldsymbol{r})]$ is a block-matrix Hamiltonian of the form\begin{align}
    \hat{\mathcal{H}}^{(\xi)}[\boldsymbol{\phi}(\boldsymbol{r})] = 
    \begin{pmatrix}
    \hat{\mathcal{H}}^{(\xi)}_{t}[\boldsymbol{\phi}(\boldsymbol{r})] & \hat{\mathcal{T}}^{(\xi)}[\boldsymbol{\phi}(\boldsymbol{r})] \\ \hat{\mathcal{T}}^{(\xi)\dagger}[\boldsymbol{\phi}(\boldsymbol{r})] & \hat{\mathcal{H}}^{(\xi)}_{b}[\boldsymbol{\phi}(\boldsymbol{r})]
    \end{pmatrix}.
\end{align}

The diagonal blocks are Dirac Hamiltonians of the top and bottom graphene layers with the account of scalar and (pseudo-)gauge potentials created by heterostrain fields, 
\begin{align}
    \hat{\mathcal{H}}^{(\xi)}_{t,b}[\boldsymbol{\phi}(\boldsymbol{r})] &=
    \hbar v_{F}\boldsymbol{\sigma}^{(\xi)}\cdot \left(-i\nabla \pm \xi \boldsymbol{\mathcal{A}}[\boldsymbol{\phi}(\boldsymbol{r})]
    \right)
    \pm \frac{\mathcal{V}}{2} \left(\partial_i\phi_i\right)\1,
\end{align}
% \begin{align}
%     \hat{\mathcal{H}}^{(\xi)}_{t,b}[\boldsymbol{\phi}(\boldsymbol{r})] &=
%     -\hbar v_{F}\boldsymbol{\sigma}^{(\xi)}\cdot \left(-i\nabla \pm \boldsymbol{\mathcal{A}}[\boldsymbol{\phi}(\boldsymbol{r})]
%     \right)
%     \pm \frac{\mathcal{V}}{2} \left(\partial_i\phi_i\right)\1,
% \end{align}
where the upper/lower sign applies to top/bottom layer block. We use the notation $\boldsymbol{\sigma}^{(\xi)} := (\xi\sigma_{x},-\sigma_{y})$ for Pauli matrices acting on the sublattice degree of freedom. The pseudogauge potential is
\begin{align}
    \boldsymbol{\mathcal{A}}[\boldsymbol{\phi}(\boldsymbol{r})] = \frac{\beta}{2a}
\begin{pmatrix} -\partial_{x}\phi_{x}+\partial_{y}\phi_{y} ,&& \partial_{x}\phi_{y}+\partial_{y}\phi_{x}\end{pmatrix}.
\end{align}
The model parameters are the Fermi velocity $v_{F} = 10^{6}$ m/s, the Grüneisen parameter $\beta=2.5$, and the bare deformation potential $\mathcal{V}= 20$ eV.
%Note that, for in-phase deformations, the expressions are completely analogous, with the caveat that the pseudomagnetic potential and strain tensor do not change sign between layers. Moreover, in that case the interlayer hopping is not affected. 

The off-diagonal blocks encode interlayer tunneling events described by\cite{castroneto2007model, bistritzer2011moire, balents2019model}
\begin{align}
    \hat{\mathcal{T}}^{(\xi)}[\boldsymbol{\phi}(\boldsymbol{r})] = \sum_{n=0,1,2} e^{i\xi (\boldsymbol{g}_{n}+\boldsymbol{Q})\boldsymbol{\phi}(\boldsymbol{r})}\hat{T}^{(\xi)}_{n}.
\end{align}
Here, $\boldsymbol{Q}$ is the position before twist of the Dirac point with chirality $\xi=+1$, and we define $\boldsymbol{g}_{0}=\boldsymbol{0}$. The matrices $\hat{T}^{(\xi)}_{n}$ are $2\pi n/3$-rotations of
$\hat{T}_{0} := w_{\textrm{AB}}\1 + w_{\textrm{AB}}\sigma_{x}$, where $w_{\textrm{AA}}$ ($w_{\textrm{AA}}$) are the intra-sublattice (inter-sublattice) tunneling rates.   %$w_{\perp}=0.0975$ eV and $w=0.3w_{\perp}$. In the absence of phonons, $\hat{H}$ reduces to the Bistritzer-MacDonald Hamiltonian.

This model evaluated at the equilibrium stacking configuration $\boldsymbol{\phi}_0(\mathbf{r})$ defines a single-electron Hamiltonian $\hat{H}_0\equiv\hat{H}[\boldsymbol{\phi}_0(\boldsymbol{r})]$ with translational symmetry in the moir\'e superlattice. Thus, it can be diagonalized in a basis of Bloch wave functions with momenta defined in the mBZ. In particular, retaining only the first term in Eq.~\eqref{eq:eq_texture}, the expressions reduce to the canonical continuum model of TBG. Displacements produced by lattice relaxation, second term in Eq.~\eqref{eq:eq_texture}, introduce higher harmonics of the moir\'e potential associated with the formation of a sharper stacking texture. In our calculations, we neglect those higher-order processes in $\hat{H}_0$ and include lattice relaxation phenomenologically by taking\cite{lu2021multiple} $w_{\textrm{AA}}=0.3\,w_{\textrm{AB}}$, with the usual value\cite{koshino2019electrons} $w_{\textrm{AB}}=0.0975$ eV.

\begin{figure}[H]
    \includegraphics[width=\linewidth]{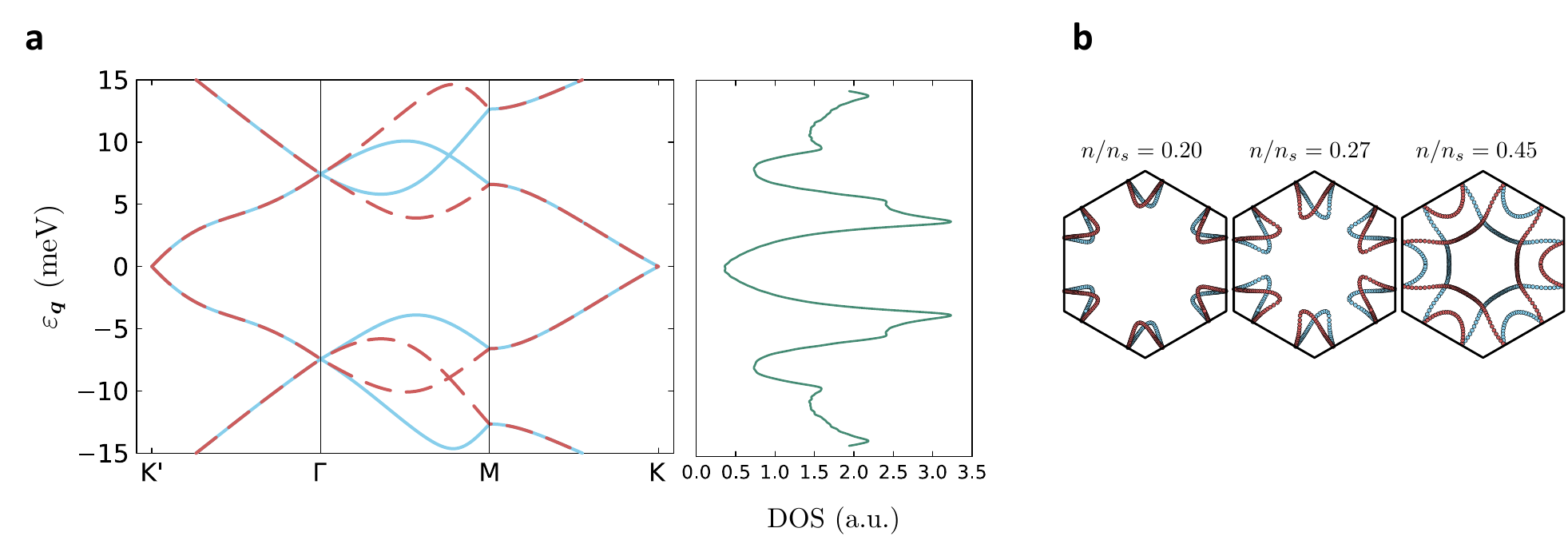}
   \caption{(a) Electron bands at $\theta = 0.4^{\circ}$, where color distinguishes valleys. The panel on the right displays the total density of states in arbitrary units.
   (b) Fermi surfaces within the mBZ at three representative fillings: below, close to, and above the Lifshitz transition.}
   \label{fig_electrons}
\end{figure}

The electronic bands deduced from this model are shown in Fig.~\ref{fig_electrons}~a. Panel~b in the same figure shows the Fermi contours corresponding to different electron fillings. The Bloch-Grüneisen temperature scale is estimated as the energy of the longitudinal acoustic fluctuation (either the phason or the in-phase phonon) evaluated at the maximum momentum transfer between electrons of the Fermi surface. Both in the case of phason and in-phase phonons the Bloch-Grüneisen temperature saturates for fillings $|n/n_{s}|\geq0.15$.

\subsubsection{Electron-phason coupling}

The electron-phason coupling follows from expanding the previous model Hamiltonian to first order in the fluctuations around the stacking texture; formally:
\begin{align}
    \hat{H}[\boldsymbol{\phi}(\boldsymbol{r})] = \hat{H}[\boldsymbol{\phi}_{0}(\boldsymbol{r}) + \delta\boldsymbol{\phi}(\boldsymbol{r})] \approx \hat{H}_0 
    + \int d\boldsymbol{r}\: \frac{\delta\hat{H}[\boldsymbol{\phi}_{0}(\boldsymbol{r})]}{\delta\boldsymbol{\phi}(\boldsymbol{r})}\delta\boldsymbol{\phi}(\boldsymbol{r})
    .
\end{align}
The first term is the electronic band Hamiltonian discussed before; the second term, $\hat{H}_{\textrm{e-ph}}$, contains the coupling of electrons with moir\'e phonons. In  our model, there are two different contributions to $\hat{H}_{\textrm{e-ph}}$: Intra-layer terms coupled to dynamical heterostrain fluctuations, and inter-layer terms representing phonon-assisted tunneling events, as recently observed with the quantum twisting microscope \cite{birkbeck2025quantum}.

To specifically isolate the contribution from phason modes to $\hat{H}_{\textrm{e-ph}}$, we expand the fluctuations in Fourier series, Eq.~\eqref{eq:fluctuation_Fourier}, and introduce the decomposition in normal modes,\begin{align}
\delta\boldsymbol{\phi}_{\boldsymbol{q}+\boldsymbol{G}}(t)=\sum_{\alpha}\delta\boldsymbol{\phi}_{\alpha,\boldsymbol{G}}(\boldsymbol{q}) \,\phi_{\alpha,\boldsymbol{q}}(t),
\end{align}
where $\delta\boldsymbol{\phi}_{\alpha,\boldsymbol{G}}(\boldsymbol{q})$ are the components of the normalized eigenvector of the dynamical matrix corresponding to mode $\alpha$. Imposing periodic boundary conditions in these coefficients, $\delta\boldsymbol{\phi}_{\alpha,\boldsymbol{G}}(\boldsymbol{q}+\boldsymbol{G}')\equiv\delta\boldsymbol{\phi}_{\alpha,\boldsymbol{G}+\boldsymbol{G}'}(\boldsymbol{q})$, hence in the normal-mode coordinates, $\phi_{\alpha,\boldsymbol{q}+\boldsymbol{G}}\equiv\phi_{\alpha,\boldsymbol{q}}$, the coupling with mode $\alpha$ in the band operator basis can be written as\begin{align}
    H_{\textrm{e-ph}}^{(\alpha)}=\frac{1}{\sqrt{A}}\sum_{\xi=\pm1}\sum_{n_1,n_2}\sum_{\boldsymbol{k}_1,\boldsymbol{k}_2\in\textrm{mBZ}}g_{\xi,n_1,n_2}^{\alpha}(\boldsymbol{k}_1,\boldsymbol{k}_2)\,\hat{c}_{\xi,n_2,\boldsymbol{k}_2}^{\dagger}\hat{c}_{\xi,n_1,\boldsymbol{k}_1}\phi_{\alpha,\boldsymbol{k}_2-\boldsymbol{k}_1},
\end{align}
where $g_{\xi,n_1,n_2}^{\alpha}(\boldsymbol{k}_1,\boldsymbol{k}_2)$ are the corresponding matrix elements of the first-order variation of the Hamiltonian with respect to normal-mode fluctuations,
\begin{align}\
    \frac{1}{\sqrt{A}}g^{\alpha}_{\xi,n_{1},n_{2}}(\boldsymbol{k}_{1},\boldsymbol{k}_{2}) = \bra{\xi,n_{2},\boldsymbol{k}_{2}} \frac{\partial\hat{\mathcal{H}}^{(\xi)}[\boldsymbol{\phi}_{0}(\boldsymbol{r})] }{\partial\phi_{\alpha,\boldsymbol{k}_{2}-\boldsymbol{k}_{1}}} \ket{\xi,n_{1},\boldsymbol{k}_{1}},
\end{align}
where $\ket{\xi,n,\boldsymbol{k}}$ is the eigenstate of $\hat{H}_0$ corresponding to an electron in valley $\xi$ and band $n$ with momentum $\boldsymbol{k}$ within the mBZ.
%Note that $\boldsymbol{k}_{2}-\boldsymbol{k}_{1}$ may lie outside of the mBZ, in which case we fold it back and include the associated umklapp phase in the coupling. We also compute the momentum-dependent Thomas-Fermi screening of the deformation potential:
The calculation of the operator between brackets follows directly from the expressions in the previous subsection after introducing the decomposition of the fluctuations in normal modes. For a doped system,  we also include the effect of screening by a Thomas-Fermi dielectric function in the intralayer scalar potential, so that the Fourier components (not restricted to mBZ) of the scalar potential read
\begin{align}
    \mathcal{V}_{\boldsymbol{k}_1,\mathbf{k}_2}=
    \frac{\mathcal{V}}{1 + \frac{e^{2}DOS(\varepsilon_{F})}{2\varepsilon_{0}|\boldsymbol{k}_1-\boldsymbol{k}_2|}},
\end{align}
where $e$ is the electron charge, $\varepsilon_{0}$ is the vacuum permittivity, and $DOS(\varepsilon_{F})$ is the electronic density of states per area at the Fermi energy.

Finally, the coupling with in-phase phonons can be calculated in the same way, with the difference that 1) intra-layer potentials enter with the same sign in both layers and 2) they do not contribute to inter-layer tunneling in our model. 

\subsubsection{Resistivity.}
The resistivity $\rho_{e-ph}$ is calculated semiclassically within Boltzmann transport theory\cite{ziman2001book}. If the stationary solutions of the linearized Boltzmann equation in the absence of temperature gradients are parametrized as $\delta f_{\boldsymbol{k}}=-\Xi_{\boldsymbol{k}}\,\partial_{\varepsilon_{\boldsymbol{k}}}n_F(\boldsymbol{k})$, where $n_F(\boldsymbol{k})$ is the equilibrium distribution given by the Fermi-Dirac function, then by the variational principle the resistivity can be found as the minimum of the following functional on $\Xi_{\boldsymbol{k}}$,
\begin{align}
    \rho_{e-ph} \leq \frac{1}{4}\frac{\frac{1}{2k_{B}T}\iint (\Xi_{\boldsymbol{k}} - \Xi_{\boldsymbol{k}'})^{2}\mathcal{P}_{\boldsymbol{k}}^{\boldsymbol{k}'} d\boldsymbol{k}d\boldsymbol{k}'}{\left|\int e\boldsymbol{v}_{\boldsymbol{k}}\Xi_{\boldsymbol{k}} \frac{\partial n_{F}(\varepsilon_{\boldsymbol{k}})}{\partial \varepsilon_{\boldsymbol{k}}} d\boldsymbol{k}\right|^{2}},
\end{align}
where the factor $1/4$ accounts for the spin and valley degeneracies (we drop the band and valley indices in the expressions of this section), $\varepsilon_{\boldsymbol{k}}$ is the electronic dispersion, $\boldsymbol{v}_{\boldsymbol{k}}$ is the electronic band velocity, and $\mathcal{P}_{\boldsymbol{k}}^{\boldsymbol{k}'}(\alpha)$ is the transition rate of the scattering by a phonon mode $\alpha$ of an electron with momentum $\boldsymbol{k}$ to a state of momentum $\boldsymbol{k}'$. Explicitly,
\begin{align}\label{eqn:Pkk}
    \mathcal{P}_{\boldsymbol{k}}^{\boldsymbol{k}'}(\alpha) &= 
    \frac{2}{\hbar} |g^{\alpha}(\boldsymbol{k},\boldsymbol{k}')|^{2} n_{F}(\boldsymbol{k})[1-n_{F}(\boldsymbol{k}')]  \text{Im}\chi_{ph}^{\alpha}(\boldsymbol{k}'-\boldsymbol{k},\varepsilon_{\boldsymbol{k}'}-\varepsilon_{\boldsymbol{k}})n_{B}(\varepsilon_{\boldsymbol{k}'}-\varepsilon_{\boldsymbol{k}}).
\end{align}
This is Fermi's Golden rule encompassing the events of emission and absorption of a phonon. On the right-hand side, $n_{B}$ is the equilibrium Bose distribution and $\chi_{ph}^{\alpha}$ is the response function of the $\alpha$-th phonon mode with dispersion $\omega_{\alpha}(\boldsymbol{k})$ and phenomenological damping coefficient $\gamma$:
\begin{align}
    \text{Im}\chi_{ph}^{\alpha}(\boldsymbol{k},\omega) := \frac{2}{\varrho} \frac{\gamma \omega}{[\omega^{2}-\omega^{2}_{\alpha}(\boldsymbol{k})]^{2}+\gamma^{2}\omega^{2}}.
\end{align}
As predicted in Ref. \cite{ochoa2023extended} , underdamped and overdamped phasons result in different $T$-dependences of the resistivity. As a phenomenological parameter, we fix the value  $\hbar\gamma = 0.6 \: k_{B}T_{\text{BG}}$ to be close to critical damping (i.e. $\hbar\gamma = k_{B}T_{\text{BG}}$, where $T_{\text{BG}}$ is the Bloch-Grüneisen temperature), so that the linear-in-$T$ regime is not extended below $T_{\text{BG}}$ and there is a transition between linear-in-$T$ and $T^{2}$ without an appreciable intermediate regime.

For a generic variational ansatz of the form $\Xi_{\boldsymbol{k}} = \sum_{n}\eta_{n}\Xi_{\boldsymbol{k}}^{(n)}$, the minimization with respect to the weights $\eta_{n}\in\mathbb{R}$ becomes a matrix inversion calculation:
\begin{align}
    \rho_{e-ph}= \frac{1}{4}\frac{1}{k_{B}T}\frac{1}{\boldsymbol{X}^{\top}P^{-1}\boldsymbol{X}},
\end{align}
where we have introduced the vector and matrix notation %(we drop the phonon index $\alpha$, which for our results is the longitudinal branch):
\begin{align}
    X_{n}:= -e\int \frac{\partial n_{F}(\varepsilon_{\boldsymbol{k}})}{\partial \varepsilon_{\boldsymbol{k}}}(\boldsymbol{v}_{\boldsymbol{k}}\boldsymbol{\hat{n}}) \Xi_{\boldsymbol{k}}^{(n)} d\boldsymbol{k},
    && 
    P_{nm} :=  \iint \Xi_{\boldsymbol{k}}^{(n)}[\Xi_{\boldsymbol{k}}^{(m)}-\Xi_{\boldsymbol{k}'}^{(m)}]\mathcal{P}_{\boldsymbol{k}}^{\boldsymbol{k}'} d\boldsymbol{k}d\boldsymbol{k}'
    ,
\end{align}
with $\boldsymbol{\hat{n}}$ being the direction of the applied electric field (without loss of generality, it can be fixed to $\boldsymbol{\hat{x}}$).
Using $\int d\boldsymbol{k} = \int d\theta d\varepsilon /|v_{\boldsymbol{k}}|$ and the quasi-elastic approximation (assuming that the temperature is much lower than the Fermi temperature), the previous expressions reduce to integrals over the Fermi surface,
\begin{align}
    X_{n} \approx
    -e \int d\theta \frac{1}{|\boldsymbol{v}^{F}_{\boldsymbol{k}_\theta}|}(\boldsymbol{v}^{F}_{\boldsymbol{k}_\theta}\boldsymbol{\hat{n}})\Xi^{(n)}_{\boldsymbol{k}_{\theta}},
\end{align}
\begin{align}
    P_{nm} &\approx
    \frac{2}{\hbar} \int \frac{d\theta d\theta'}{|\boldsymbol{v}_{\boldsymbol{k}_\theta}^{F}||\boldsymbol{v}_{\boldsymbol{k}_\theta'}^{F}|} \Xi_{\boldsymbol{k}_{\theta}}^{(n)}[\Xi_{\boldsymbol{k}_{\theta}}^{(m)}-\Xi_{\boldsymbol{k}_{\theta'}'}^{(m)}] |g^{\alpha}(\boldsymbol{k}_\theta,\boldsymbol{k}_{\theta'}')|^{2} 
    \int d\omega\: \omega\: \text{Im}\chi_{ph}^{\alpha}(\boldsymbol{k}_{\theta'}'-\boldsymbol{k}_{\theta},\omega)n_{B}(\omega)[1+n_{B}(\omega)].
\end{align}
The last integral over frequencies can be solved analytically as in Ref. \cite{ochoa2023extended} . The decay of the Bose distribution and phonon response function allows one to calculate it over the whole real line, instead of over a small interval around $\omega=0$.

\subsection{Numerical implementation of the model.}
To produce the dispersions of phonons we use 55 stars in our codes, and 15 stars to calculate the electron bands, wavefunctions, and matrix elements of the electron-phonon coupling. Each star is a collection of 6 moir\'e reciprocal lattice vectors related by $C_{6}$ rotations. In the calculation of the density of states, we include a factor of 2 to account for the spin degeneracy and a broadening $\zeta = 0.3$ meV:
\begin{align}
    DOS(\omega) = \frac{2}{A}\sum_{\boldsymbol{q}}^{\text{mBZ}}\sum_{\xi}\sum_{n}^{\text{bands}} \frac{\zeta}{(\hbar\omega - \varepsilon_{n,\xi,\boldsymbol{q}})^{2} + \zeta^{2}}.
\end{align}

In the estimation of the resistivity, we compute integrals over Fermi surfaces, so we choose our variational ansatz adapted to its geometry. 
%We do this in two different ways, since the Fermi surface is disconnected (connected) for fillings below (above) the Lifshitz transition. 
For a connected Fermi surface (for fillings beyond the Lifshitz transition), we went beyond the relaxation-time approximation by introducing a variational ansatz expanded in a basis of cylindrical harmonics: $\Xi_{\boldsymbol{k}}^{(n)}= \Xi_{\theta}^{(n)}:=  \cos(n\theta)$. 
The contribution from odd terms would vanish because of the $D_{3}$ symmetry of each valley.
Naively, the more elements we include in our variational basis, the more accurate the calculation is; but numerically we are restricted by the number of points in the discretization of the Fermi contours. %Nyquist theorem. 
This was considered when we checked the convergence of the results with respect to the number of terms included in the ansatz.

In the case of a disconnected Fermi surface, the previous variational guess is not suitable because its efficiency relies on the orthogonality of different harmonics, yet the support of the integrands in that case is only a subset of $[0,2\pi)$ and therefore the restrictions of the functions $\Xi^{(n)}_{\theta}$ are no longer orthogonal. 
The smaller the Fermi pockets (i.e. the closer to charge neutrality), the worse the cylindrical harmonics perform as an ansatz. In those cases, we computed the resitivity in the relaxation-time approximation, $\Xi_{\boldsymbol{k}}= \boldsymbol{\hat{n}}\cdot \nabla_{\boldsymbol{k}} \varepsilon_{\boldsymbol{k}}$.
%A Fermi surface is the surface level $\{\boldsymbol{k}\:|\:\varepsilon_{\boldsymbol{k}}=\varepsilon_{F}\}$, hence its geometry is given by the gradient of the energy dispersion $\nabla_{\boldsymbol{k}} \varepsilon_{\boldsymbol{k}} = \boldsymbol{v}_{\boldsymbol{k}}$.
%We define the trial function $\Xi_{\boldsymbol{k}}= \boldsymbol{v}_{\boldsymbol{k}}\boldsymbol{\hat{n}}$. 
As there is not variational parameter in this case, we test this second approach on connected Fermi surfaces and compared the results with the preceding calculations, obtaining very similar results.

Because of the computational time needed to calculate all-to-all scatterings, we introduce the additional approximation $g(\boldsymbol{k}_\theta,\boldsymbol{k}_\theta')\approx g(|\boldsymbol{k}_\theta-\boldsymbol{k}_{\theta'}'|)$, which is exact only for spherical Fermi surfaces.
In addition, the deformation of the Fermi pockets around the Lifshitz transition shown in  Fig.\ref{fig_electrons}(b) makes the angular integration ill-defined as the integrand is not uniquely-valued at some  $\theta$. This is the reason why we do not show theoretical results for those fillings.

\subsection{Weak insulating behavior at full band filling}

\begin{figure}[H]
    \includegraphics[width=\linewidth]{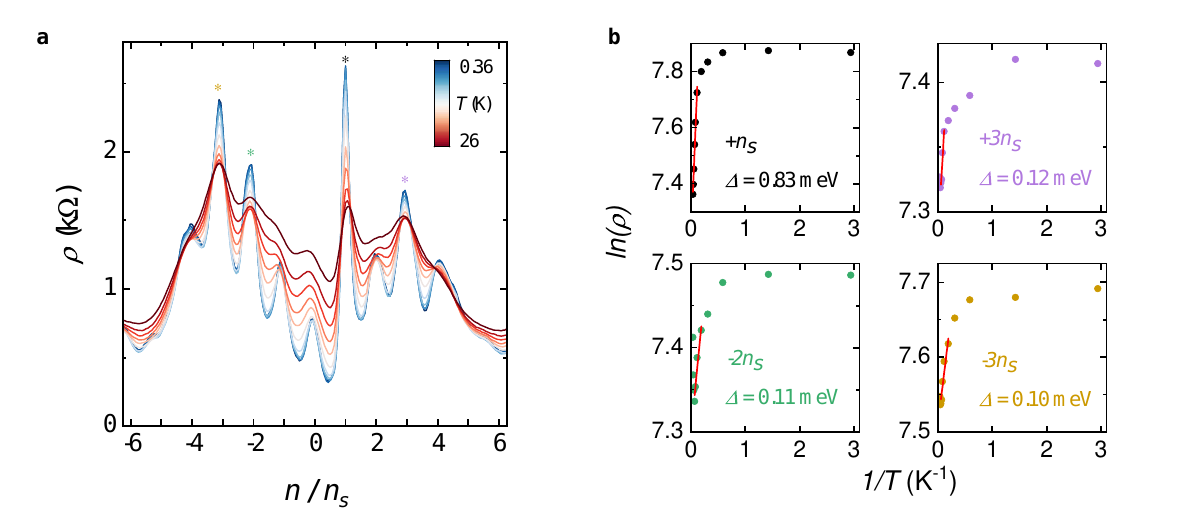}
   \caption{(a) Low-temperature resistivity curves as a function of band filling. The insulating peaks are marked with an asterisk. (b) Natural logarithm of the resistivity as a function of inverse temperature for the peaks marked in panel a. Solid red lines are fits to $\rho \propto exp{[\Delta/(2k_BT)]}$, where $k_B$ is the Boltzmann constant and $\Delta$ is the energy gap (fitted values are indicated).}
   \label{figSI_insulating}
\end{figure}

At several full filling conditions, we find that the low-temperature resistivity of mTBG decreases as temperature increases, as shown in Figure \ref{figSI_insulating}a. These insulating peaks push the onset of the linear-in-$T$ behavior to slightly higher temperature with respect to partial filling regions. The activated temperature dependence, shown in Figure \ref{figSI_insulating}b, suggests that small energy gaps separate some of the moiré bands. Contrary to Ref. \cite{lu2021multiple} , the insulating trend is observed in absence of a perpendicular magnetic field.

\subsection{(Magneto)transport and linear-in-$T$ coefficient from additional contact configurations}

\begin{figure}[H]
    \includegraphics[width=\linewidth]{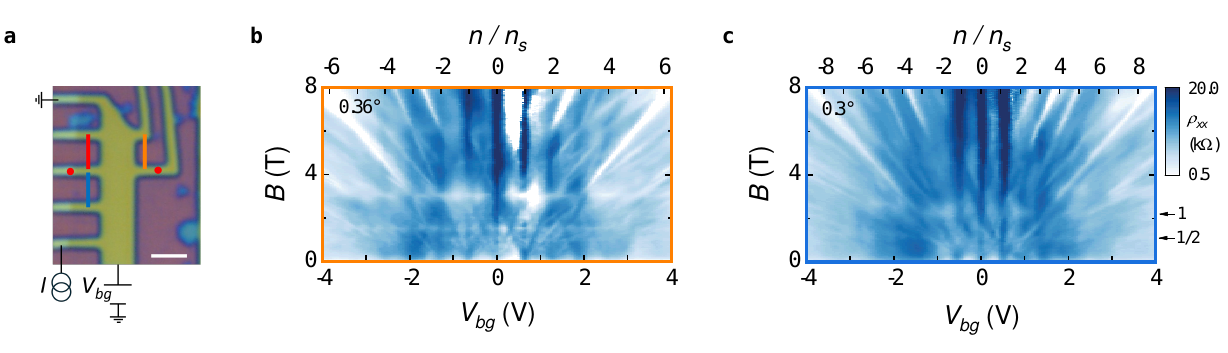}
   \caption{(a) Optical microscopy image of the device, with  measurement configuration  sketched. The red line indicates the voltage probes used for resistivity measurements presented in the main text; the two red dots show the contacts used for Hall effect measurements. Additional longitudinal voltage probes investigated are indicated by the orange and light blue line. The white scale bar is 2 $\mu$m. (b-c) Longitudinal resistivity as a function of gate voltage and magnetic field,  measured at $T=0.36$ K on additional voltage probes (the frame color indicates the corresponding contacts in panel a). On the top axis, the band filling is indicated. The estimated twist angle is reported on top left corner. The same color scale is used for the two panels.}
   \label{figSI_transpHof}
\end{figure}

We test the reproducibility of the device response by measuring the longitudinal resistivity with alternative voltage probes. The contacts used for data reported in the main text are indicated by the red line (longitudinal) and the red dots (Hall) in Figure \ref{figSI_transpHof}a. Employing the longitudinal contact pair on the opposite side of the mesa (orange line), we measure an Hofstader’s butterfly  that quantitatively reproduces the features discussed in the main text (Figure \ref{figSI_transpHof}b; $n_s=0.32\times10^{12}$ cm$^{-2}$, $\phi/\phi_0=1$ at $B=3.1$ T). This indicates that the upper region of the device hosts a uniform moiré pattern. In the lower part of the sample, as probed by the blue contact pair, we obtain an Hofstader’s butterfly dominated by a smaller twist angle. Based on $n_s=0.21\times10^{12}$ cm$^{-2}$, $\phi/\phi_0=1$ at $B=2.2$ T, we estimate $\theta=0.3^\circ$. Sub-$0.1^\circ$ angle variations over few micrometers are typical in TBG devices \cite{uri2020mapping}. The minimal twist angle employed in this experiment amplifies the visibility of these variations, since they correspond to large changes in the moiré wavelength ($\lambda=39$ nm to 47 nm across our device; for comparison, the same variation in $\theta$ around magic angle would correspond to a change in $\lambda$ of only 0.7 nm). We further note that the data acquired in the lower area show generally broader resistivity features, suggesting that a major twist angle variation takes place in the region within the blue contact pair, however not affecting the upper part of the sample.

\begin{figure}[H]
    \includegraphics[width=\linewidth]{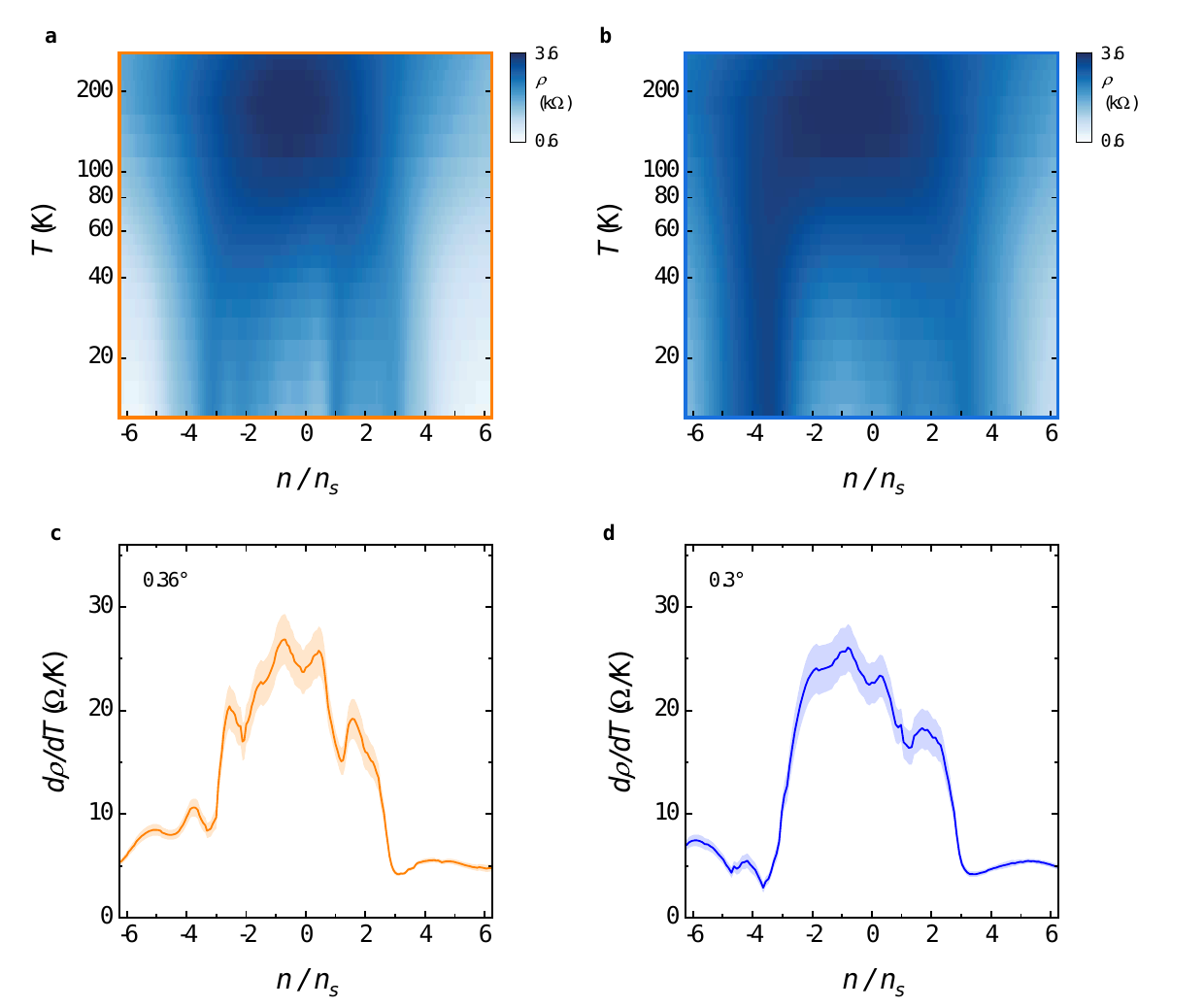}\caption{(a,b) Resistivity maps as a function of band filling and temperature (the frame color indicates the corresponding contacts in Figure \ref{figSI_transpHof}a) (c,d) Extracted $d\rho/dT$ from linear fits, plotted as function of band filling. The shaded red area corresponds to $\pm$ one standard error from the fits.}
   \label{figSI_transpTdep}
\end{figure}

We use the additional voltage probes also to measure $\rho(T)$. Color maps of the resistivity as a function of band filling and temperature are shown in Figure \ref{figSI_transpTdep}a-b. We note that the $n/n_s$ axis normalizes the difference in twist angle between the two regions. The dome-like feature discussed in the main text, bounded by $T\sim80$ K and $n/n_s=\pm3$ is reproduced by both measurement configurations. The $0.3^\circ$ area shows less resolved resistivity peaks, especially for hole doping. Both data sets include an extended region of linear-in-$T$ resistivity, which is fitted following the protocols presented in the next Section. The resulting slope $d\rho/dT$ is shown in Figure \ref{figSI_transpTdep}c-d. In both cases $d\rho/dT$ is stronger within $|n/n_s|<3$, reaching the same order  of magnitude discussed in the main text. The modulation of $d\rho/dT$ visible in main text Figure 3a is reproduced by the orange contact pair, while it is less defined in the $0.3^\circ$ region (blue contact pair). While a device-scale twist angle variation is typically undesirable, in this case it suggests that the phenomenology presented in this study is not limited to the specific angle discussed in the main text, but rather a general feature of mTBG.

\subsection{Resistivity in the intermediate and high temperature ranges}

\begin{figure}[H]
    \includegraphics[width=\linewidth]{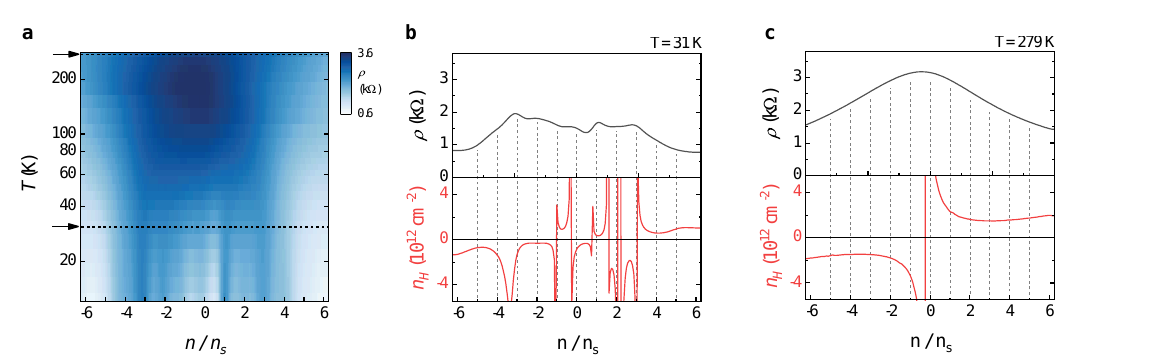}
   \caption{(a) Resistivity of mTBG as a function of band filling and temperature (Log scale), same plot of Fig. 3a in the main text. Dashed lines and arrows indicate the $\rho(n/n_{s})$ profiles in panels b,c. Resistivity (black line, top panel) and Hall carrier density (red line, bottom panel) as a function of  band filling at (b) $T=31$ K and (c) $T=297$ K. }
   \label{figSI_add_roxx_cuts}
\end{figure}

We present the zero-field resistivity $\rho$ and the Hall carrier density $n_{H}$ (at $B=0.25$ T) as functions of band fillings for two temperatures representative of the intermediate (Fig. \ref{figSI_add_roxx_cuts}b) and high (Fig. \ref{figSI_add_roxx_cuts}c) ranges discussed in the main text.
At $T=31$ K, $\rho$ exhibits an intensity modulation near integer band filling values, which is accompanied by multiple features in $n_{H}$. In contrast,  at $T=297$ K,  $\rho$ presents a single broad peak and $n_{H}$ shows a single sign change at zero filling.

\subsection{Fitting protocols for $\rho(T)$}

We employ the following fitting protocol to obtain the linear coefficient of longitudinal resistivity as a function of temperature at each band filling value.
\begin{enumerate}[itemsep=2pt, topsep=3pt, label=\roman*.]
    \item we use a first order polynomial $\rho(T)=(d\rho/dT) \cdot T + c$ as a fitting function, where $(d\rho/dT)$ and $c$ are the fit coefficients; 
    \item for each value of $n/n_s$, we least squares minimize the experimental curve $\rho(T)$ in all temperature intervals with at least 6 data points and we estimate the resulting reduced chi-squared coefficients; 
    \item we exclude all the fits with reduced chi-squared coefficient less than 0.998; 
    \item from the remaining set, we select the fits that maximize the number of experimental data points to obtain the $d\rho/dT$ curves, which are shown in main text Figure 3e and Figure \ref{figSI_transpTdep}c-d.
\end{enumerate}
We cross-check the results by fitting all experimental curves in a fixed temperature interval: 26 K - 66 K. This second method gives comparable values of $d\rho/dT$ as a function of band filling. \\
The quadratic fits of $\Delta\rho(T)$ shown in Figure 4 of the main text are calculated following a similar protocol. \\
% \begin{enumerate}[itemsep=2pt, topsep=3pt, label=\roman*.]
    - we re-scale the resistivity by subtracting the value at base temperature ($T=0.36$ K) to isolate the temperature-driven component $\Delta\rho$; \\
    - for each value of $n/n_s$, we linearly least square fit the low-temperature data set to  $\Delta\rho(T)=A \cdot T^2$ in all possible temperature ranges starting from 0.7 K and including at least 5 data points; \\
    - we select the fit in the largest temperature range and having a reduced chi square higher than 0.999.

\subsection{Linear-in-$T$ and $T^2$ behavior across different moiré bands}

\begin{figure}[H]
    \includegraphics[width=0.95\linewidth]{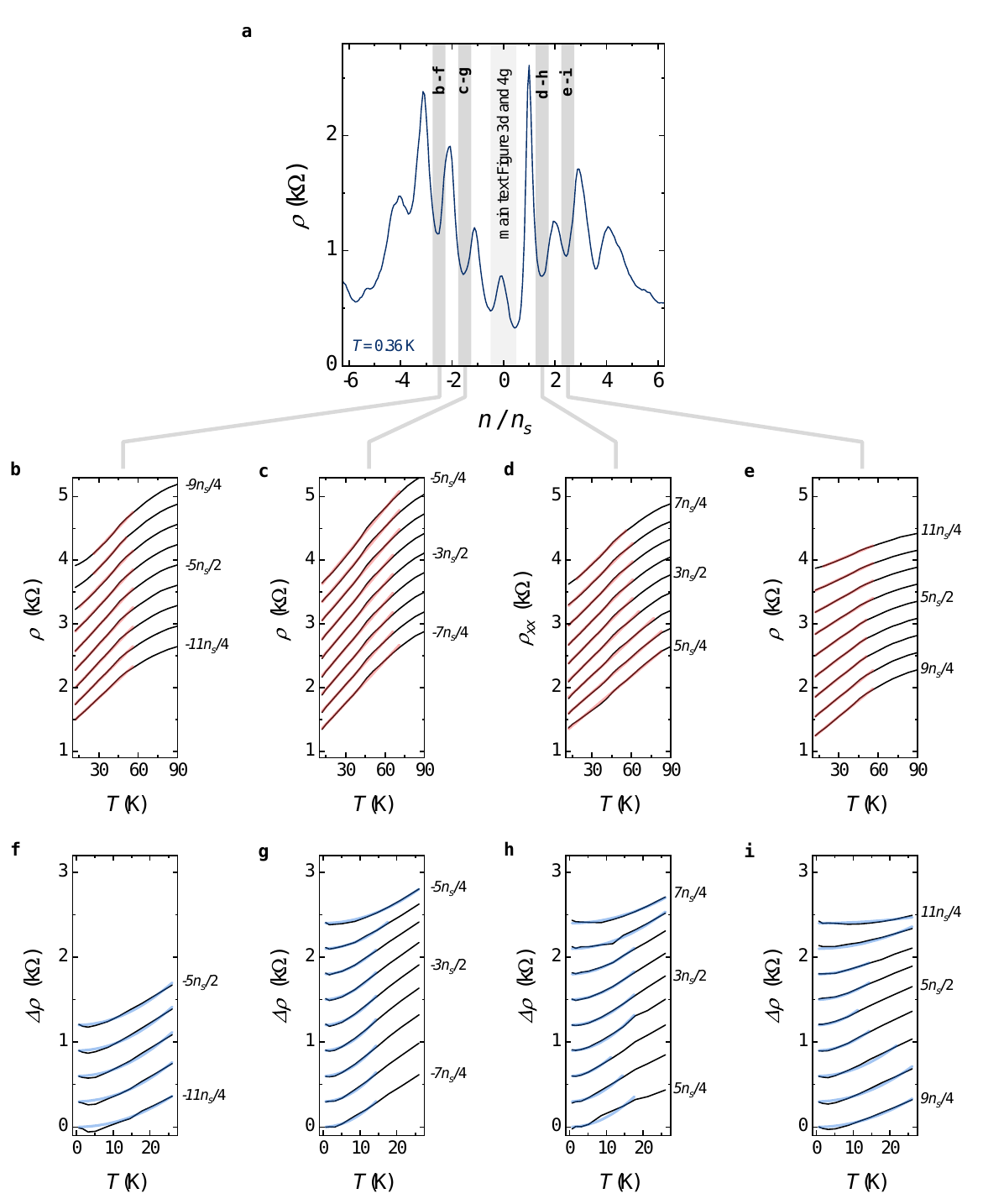}
   \caption{\footnotesize (a) $\rho$ measured at base temperature ($T=0.36$ K) as a function of band filling. The shaded areas indicate the density ranges considered in the other panels. (b-e) $\rho(T)$ at intermediate temperatures (black) and linear fits (red), measured within the third hole (b), second hole (c), second electron (d), and third electron (e) band. (f-i) $\Delta\rho(T)$ at low temperatures (black) and quadratic fits (blue), measured within the third hole (f), second hole (g), second electron (h), and third electron (i) band. The curves in panels b-i are offset by 0.3 k$\Omega$ for clarity.}
   \label{fig_TandT2}
\end{figure}

We present additional temperature-dependent data that highlight the identified trends across the moiré bands. Figure \ref{fig_TandT2}a shows $\rho$ as a function of $n/n_s$, serving as a reference for the filling of moiré bands, and shaded areas corresponding to the ranges considered in the different panels. Data plotted in main text Figure 3d and 4g are from the first electron and hole bands. In Figure \ref{fig_TandT2}b-e we show representative $\rho$ curves from the second and third electron and hole bands in the linear-in-$T$ regime, with linear fits superimposed. Figure \ref{fig_TandT2}f-i show $\Delta\rho$ in the $T^2$ regime, along with quadratic fits. We exclude data in the vicinity of the insulating peaks, where negative $\Delta\rho$ are obtained at low temperature; this characteristic is found to extend over part of the third hole band (Figure \ref{fig_TandT2}f).

\end{bibunit}

\end{document}